\documentclass[aps,prd,reprint,groupedaddress]{revtex4-1}
\usepackage{amsmath}
\usepackage{amsfonts}
\usepackage{amssymb}
\usepackage{graphicx}

\newcommand{\etal}{\textit{et al}.}

\begin{document}

\title{Covariant Effective Action for Antisymmetric Tensor Field}%

\author{Sandeep Aashish}%
\email[]{sandeepa16@iiserb.ac.in}

\author{Sukanta Panda}
\email[]{sukanta@iiserb.ac.in}
\affiliation{Department of Physics, Indian Institute of Science Education and Research, Bhopal - 462066, India}

\date{\today}

\begin{abstract}
Covariant quantization of rank-2 antisymmetric fields is non-trivial due to additional symmetries of the gauge parameters. We present an intuitive way to deal with this additional symmetry of gauge parameters in terms of geometrical understanding of field space. We generalize the DeWitt-Vilkovisky covariant effective action formalism for quantization of such theories. As an application, we quantize a massive rank-2 antisymmetric field using the covariant effective action approach and reproduce previously obtained results.  
\end{abstract}

\maketitle

\section{Introduction}
The significance of antisymmetric tensor fields in the context of superstring theories is well established\cite{rohm1986,buchbinder2008}. Rank-2 antisymmetric fields with minimal and non-minimal potential terms have also been considered by Altschul \etal \cite{altschul2010} to study spontaneous Lorentz violation\cite{kostelecky2004}. A strong motivation for such a study is the possibility of experimental signals for Lorentz violation involving gravitational couplings\cite{altschul2010}. A natural extension of this endeavour is to study quantum aspects of antisymmetric fields. One of the first works in this direction is undertaken in Buchbinder \etal\cite{buchbinder2008} where quantization of massive rank-2 and rank-3 antisymmetric models (minimal model with potential being the mass term) is discussed to study their quantum equivalence to massive vector and scalar fields respectively. Even before, some aspects of quantization of antisymmetric fields have been discussed in \cite{siegel1980,hata1981,buchbinder1988,duff1980,bastianelli2005a,bastianelli2005b}.

However, previous studies have not employed a generally covariant formalism for study of quantum corrections to antisymmetric fields. Recently, a covariant effective action approach was used by Saltas and Vitagliano\cite{saltas2017} to write the 1-loop effective action for a cubic Galileon model including quantum gravitational corrections. The covariant effective action approach, introduced by DeWitt-Vilkovisky, is an efficient tool to study effective action in a way that is gauge invariant and background field invariant\cite{dewitt1964,dewitt1967a,dewitt1967b,dewitt1967c,vilkovisky1984a,vilkovisky1984b}. Moreover, features such as frame independence can be introduced in this formalism by taking into account the conformal transformations in addition to field reparametrizations\cite{steinwachs2013,steinwachs2015,karamitsos2018,*karamitsos2017}, making it an ideal tool to study quantum gravitational effects in the context of spontaneously Lorentz-violating models of antisymmetric fields, and other cosmological models in general. 

Unlike simple rank-1 antisymmetric (electromagnetic) fields, quantization of rank-2 or higher antisymmetric fields is non-trivial especially because of the additional symmetries of gauge parameters of the theory\cite{buchbinder2007}. A simple, ad hoc resolution applicable to the case of antisymmetric fields was discussed by Buchbinder and Kuzenko\citep{buchbinder1988}. More general but complex resolutions to this problem have been discussed before in literature\cite{schwarz1978,schwarz1979,batalin1983}. Moreover, quantization of massive antisymmetric models has an additional challange, as these models suffer from softly broken gauge symmetries, and require the use of St{\"u}ckelberg procedure\cite{stuckelberg1957} to restore softly broken gauge freedom before quantizing the theory. 

In the present work, we use DeWitt-Vilkovisky's geometrical understanding of field space and gauge fixing to generalize the covariant effective action formalism for quantizing massive rank-2 antisymmetric fields. We give an intuitive, geometric prescription for dealing with symmetries of gauge parameters while quantizing the theory. We then write the covariant effective action for a massive rank-2 antisymmetric field. In an attempt to be pedagogical, major steps leading to the effective action have also been presented.
In essence, this paper lays the foundation for future studies involving the quantum corrections in the context of rank-2 antisymmetric fields with spontaneously broken Lorentz symmetry, in a covariant language. 

The organization of this paper is as follows. In section 2, we describe the set-up of our problem, including the geometrical notations used and the action for the antisymmetric field. Section 3 is devoted to generalizing quantization of theories with gauge parameters having additional symmetries. We then proceed with the calculation of the covariant effective action in section 4.

\section{\label{sec2} The Set-up}
In our calculations we follow the general procedure of the book by Toms and Parker\cite{parker2009}. This section will be devoted to briefly reviewing the geometric or condensed notations introduced by DeWitt\cite{dewitt1964} and the steps leading to the covariant effective action. We then introduce the action of the rank-2 antisymmetric field to be quantized. 

\subsection{Geometric notations}
Effective action $\Gamma[\bar{\phi}]$ is defined as\footnote{For a formal definition of effective action, refer standard textbooks in QFT, as well as ref. \cite{parker2009} and \cite{buchbinder1992}}
\begin{equation}
\label{geo1}
\exp\left(\dfrac{i}{\hbar}\Gamma [\bar{\phi}]\right)= \mathcal{N}\int D\phi\exp\left\{\dfrac{i}{\hbar}S[\phi ]+\dfrac{i}{\hbar}(\phi-\bar{\phi})\Gamma_{,\phi}[\bar{\phi}]\right\}
\end{equation}
The above expression is not generally invariant under coordinate transformations and field re-parametrizations, both of which are natural requirements for a physical theory. Moreover, the treatment of gauge theories using the Faddeev-Popov method in (\ref{geo1}) leads to a background and gauge condition dependent effective action\cite{buchbinder1992}. 
A covariant effective action, free of gauge and background dependence was achieved by DeWitt\cite{dewitt1964,dewitt1967a,dewitt1967b,dewitt1967c} and Vilkovisky\cite{vilkovisky1984a,vilkovisky1984b}.

Invariance of the effective action under coordinate transformations and field redefinitions is realised by going to the space of fields with field components as the coordinates in field space. In DeWitt's condensed notation, field components are denoted by vector components $\varphi^{i}$ with index $i$ mapped to tensor as well as spacetime indices of field components. 
All the field components (variables) in the action are represented by $\varphi^{i}$. For example, if a field variable $A_{\mu}(x)$ is denoted by $\varphi^{i}$ in field space, then $i$ is mapped to both the tensor index and coordinate index i.e. $i\longrightarrow (\mu,x)$. The summation convention still follows here, which implies that a sum over field space indices will correspond to sum over vector indices and integral over spacetime indices, i.e.
\begin{equation}
\label{geo2}
g_{ij}v^{i}w^{j} = \int d^{n}x d^{n}x' g_{IJ}(x,x')v^{I}(x)w^{J}(x').
\end{equation}
The Dirac $\delta$-distribution in field space is defined as 
\begin{equation}
\label{geo3}
\delta^{i}_{j} = |g(x')|^{1/2}\delta^{I}_{J}\delta(x,x') \equiv \delta^{I}_{J} \tilde{\delta}(x,x'),
\end{equation}
where, $\tilde{\delta}(x,x')$ transforms as a scalar in first argument, and scalar density in second argument. This definition of $\tilde{\delta}(x,x')$ carries over to the field space Dirac $\delta$-distribution as well. 
The functional derivative is given by
\begin{eqnarray}
\label{geo4}
\varphi^{i}_{,j} \equiv \dfrac{\delta\varphi^{I}(x)}{\delta\varphi^{J}(x')} = |g(x')|^{1/2}\delta^{I}_{J}\delta(x,x') = \delta^{i}_{j}.
\end{eqnarray}
A metric $g_{ij}$ can be defined in the field space with properties analogous to the spacetime metric: $g_{ij}g^{jk} = \delta_{i}^{k}$, along with the Christoffel connections 
\begin{equation}
\label{geo5}
\Gamma^{k}_{ij} = \frac{1}{2}g^{kl}(g_{il,j} + g_{lj,i}-g_{ij,l}).
\end{equation}
Since we will be dealing with covariant quantities in our calculations, we will denote the invariant volume element $\sqrt{-g(x)}d^{n}x$ with $dv_{x}$ henceforth ($g(x)$ is the spacetime metric).
A gauge transformation is given by 
\begin{eqnarray}
\delta\varphi^{i} = K^{i}_{\alpha}[\varphi]\delta\epsilon^{\alpha},
\end{eqnarray}
where $\epsilon^{\alpha}$ is the gauge parameter and $K^{i}_{\alpha}$ are generators of gauge transformation. For a covariant field-space calculation, one needs to use covariant intervals $\sigma^{i}[\varphi_{*};\varphi]$, which are a generalization of flat field space intervals $\varphi^{i} - \varphi^{i}_{*}$ (where $\varphi_{*}^{i}$ are fixed points in field space) and are defined as,\begin{eqnarray}
\sigma^{i}[\varphi_{*};\varphi] = g^{ij}\dfrac{\delta}{\delta\varphi^{j}}\sigma[\varphi_{*};\varphi],
\end{eqnarray}
where $\sigma[\varphi_{*};\varphi]$ is the geodetic interval defined as,
\[
\sigma[\varphi_{*};\varphi] = \dfrac{1}{2}(length \ of \ geodesic \ from \ \varphi^{i}_{*} \ to \ \varphi^{i})^{2}.
\]

\subsection{Action for the free antisymmetric rank-2 tensor field}
In order to set up a covariant formalism to study quantum corrections to models with antisymmetric fields, it is relevant to consider action for the minimal model discussed by Altschul \etal \cite{altschul2010}, with potential being the mass term,
\begin{equation}
\label{act1}
S[B] = \int d v_{x} \left\{-\dfrac{1}{12}F^{\mu\nu\lambda}[B]F_{\mu\nu\lambda}[B] - \dfrac{1}{4}m^{2}B^{\mu\nu}B_{\mu\nu}\right\},
\end{equation}
where, 
\begin{equation}
\label{act2}
F_{\mu\nu\lambda}[B] \equiv \nabla_{\mu}B_{\nu\lambda} + \nabla_{\lambda}B_{\mu\nu} + \nabla_{\nu}B_{\lambda\mu}.
\end{equation}
The action (\ref{act1}) belongs to the class of theories with softly-broken gauge symmetry\cite{buchbinder2007}, as the kinetic term is invariant under the transformation
\begin{equation}
\label{act3}
B_{\mu\nu}\longrightarrow B^{\xi}_{\mu\nu} = B_{\mu\nu} + \nabla_{\mu}\xi_{\nu}-\nabla_{\nu}\xi_{\mu},
\end{equation}
while the mass term $m^{2}B^{\mu\nu}B_{\mu\nu}/4$ is not. It has been shown that, such theories contain the redundant degrees of freedom but cannot be dealt with using traditional Faddeev-Popov method\cite{buchbinder2007,barvinsky1985}. A convenient way to deal with this problem is to restore the softly broken symmetry\cite{buchbinder2007} of the theory using St\"{u}ckelberg procedure\cite{stuckelberg1957}. One introduces a new field $C_{\mu}$ such that, 
\begin{eqnarray}
\label{act4}
S[B,C] = \int d v_{x} \left\{-\dfrac{1}{12}F^{\mu\nu\lambda}[B]F_{\mu\nu\lambda}[B] - \right. \nonumber \\ \left.\dfrac{1}{4}m^{2}(B^{\mu\nu} + \frac{1}{m}H^{\mu\nu}[C])^{2}\right\},
\end{eqnarray}
where, $H_{\mu\nu}[C]\equiv \nabla_{\mu}C_{\nu}-\nabla_{\nu}C_{\mu}$. The new action (\ref{act4}) has the following symmetries:
\begin{eqnarray}
\label{act5}
B_{\mu\nu}&\longrightarrow & B^{\xi}_{\mu\nu} = B_{\mu\nu} + \nabla_{\mu}\xi_{\nu}-\nabla_{\nu}\xi_{\mu}, \nonumber \\
C_{\mu}&\longrightarrow & C^{\xi}_{\mu} = C_{\mu} -m\xi_{\mu},
\end{eqnarray}
and 
\begin{eqnarray}
\label{act6}
C_{\mu}&\longrightarrow & C^{\Lambda}_{\mu} = C_{\mu} + \nabla_{\mu}\Lambda, \nonumber \\
B_{\mu\nu}&\longrightarrow & B^{\Lambda}_{\mu\nu} = B_{\mu\nu},
\end{eqnarray}
and reduces to the original theory (\ref{act1}) in the gauge $C_{\mu}=0$. Since our approach is gauge invariant, we can work with the full theory (\ref{act4}) instead of (\ref{act1}) and choose any suitable gauge condition. 
As is encountered later, particular choices of gauge condition lead to further softly broken symmetry in the St\"{u}ckelberg field, and successive application of St\"{u}ckelberg procedure is the key to resolving such cases.\\
The theory (\ref{act4}) is, however, still not free from degeneracies because of the extra symmetry of the gauge parameter $\xi_{\mu}$, 
\begin{equation}
\label{act7}
\xi_{\mu} \longrightarrow \xi^{\psi}_{\mu} = \xi_{\mu} + \nabla_{\mu}\psi, \quad
\Lambda \longrightarrow \Lambda + m\psi .
\end{equation} 
leaving fields $B_{\mu\nu}$, $C_{\mu}$ invariant. We give a geometrical prescription for dealing with this issue and generalize the quantization of such theory in the next section.

\section{Dealing with the symmetries of gauge parameters}
In the field space, set of points $\{\varphi^{i}_{\epsilon}\}$ connected by the gauge parameter $\epsilon^{\alpha}$ form an orbit called gauge orbit. So, fixing a gauge is equivalent to selecting one point from each gauge orbit. This is achieved by setting up a coordinate system $(\xi^{A},\theta^{\alpha})$ such that coordinates $\theta^{\alpha}$ are along the orbit (longitudinal) while coordinates $\xi^{A}$ are transverse to the orbit. Fixing $\theta^{\alpha}$ is then equivalent to choosing one point from the orbit. Gauge invariant quantities are defined as having no $\theta^{\alpha}$ dependence. As is standard practice, we assign $\theta^{\alpha} = \chi^{\alpha}[\varphi]$, where $\chi^{\alpha}[\varphi]$ is the gauge-fixing condition for fields $\varphi^{i}$.

In the case of (\ref{act4}), however, the gauge parameters $\xi^{\mu}$ too have symmetry given by (\ref{act7}). This means, for every choice of $\theta^{\alpha}$ there exists an equivalence class (a set of points $\{\xi^{\psi}_{\mu}\}$) in the space of gauge parameter $\xi_{\mu}$. To deal with this issue, we follow the familiar procedure of `fixing the gauge' in parameter space. What this means in the geometric picture is as follows. We will work in condensed notation for this purpose. 

\begin{figure}
\includegraphics[scale=0.35]{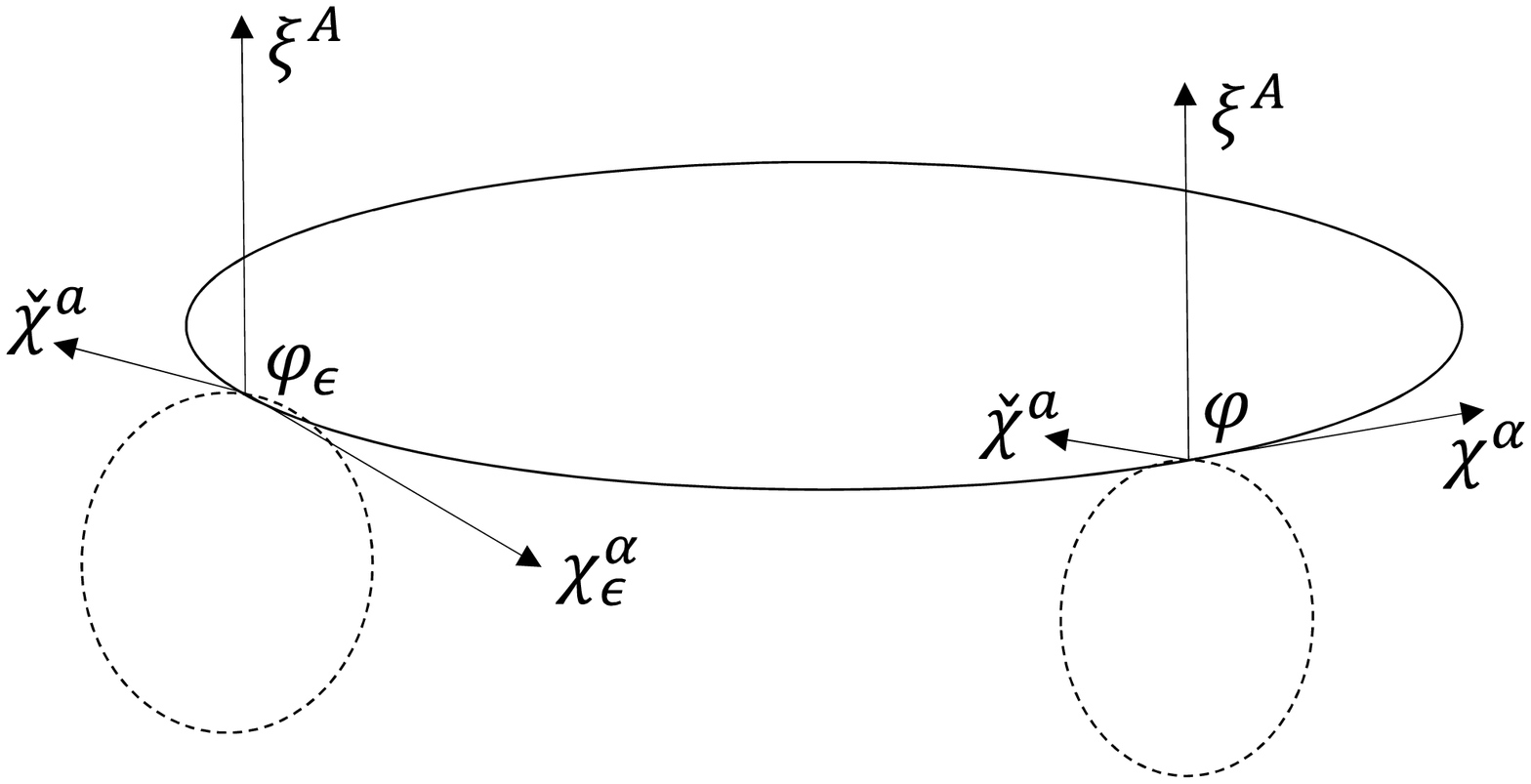}
\caption{\label{fig1} An illustration of gauge orbit in field space (solid line) and parameter space orbit (dashed line). At different points on gauge orbit ($\varphi$ and $\varphi_{\epsilon}$ respectively), a point $\check{\chi}^{a}$ on parameter-space orbit remains fixed.}
\end{figure}

Let us denote gauge parameters by $\epsilon^{\alpha}$ where $\alpha$ is a condensed index mapped to $(\mu,x)$. We are interested in the case where $\epsilon^{\alpha}$ has a gauge freedom that leaves $\varphi^{i}$ unchanged, having a general form
\begin{eqnarray}
\label{dgp1}
\delta\epsilon^{\alpha} = \check{K}^{\alpha}_{a}[\epsilon]\delta\lambda^{a},
\end{eqnarray}
where $\lambda^{a}$ parametrizes the transformations of $\epsilon^{\alpha}$. It is assumed that $\lambda^{a}$ are free of any such symmetry. 
The usual Faddeev-Popov method of gauge fixing involves introducing a factor, 
\begin{eqnarray}
\label{dgpn1}
1 = \int \left(\prod_{\alpha}d\chi^{\alpha}\right)\tilde{\delta}[\chi^{\alpha}[\varphi_{\epsilon}];0] ,
\end{eqnarray}
in the path integral, to calculate an appropriate gauge-fixed measure. However, in the present case, a technical difficulty with (\ref{dgpn1}) is that the measure spans all $\epsilon^{\alpha}$ including points on parameter-space orbit ($\epsilon^{\alpha}_{\lambda}$).
In order to deal with this issue, we start by revisiting the condition for gauge fixing in field space, that is, the requirement for $\chi^{\alpha}[\varphi]$ to be a gauge-fixing condition. $\chi^{\alpha}[\varphi]$ is required to be unique at each point $\varphi_{\epsilon}$ on a gauge orbit. This translates to requiring that the equation
\begin{equation}
\label{dgp2}
\chi^{\alpha}[\varphi_{\epsilon}] = \chi^{\alpha}[\varphi],
\end{equation} 
have a unique solution $\delta\epsilon^{\alpha} = 0$. Expanding left hand side about $\varphi$ yields, 
\begin{equation}
\label{dgpr1}
Q^{\alpha}_{\beta}[\varphi]\delta\epsilon^{\beta} = 0,
\end{equation}
where, $Q^{\alpha}_{\beta}[\varphi] = \chi^{\alpha}_{,i}[\varphi]K^{i}_{\beta}[\varphi]$. The condition for unique solution is $\det{Q^{\alpha}_{\beta}}\neq 0$. But, it turns out that for theories with degeneracy in the gauge parameter (of the form eq. (\ref{dgp1})), determinant of $Q^{\alpha}_{\beta}$ vanishes, making condition (\ref{dgp2}) insufficient for gauge fixing in this case\cite{buchbinder1992}. Indeed, for theory (\ref{act4}), it can be explicitly checked that (\ref{dgp1}) is a solution to eq. (\ref{dgpr1})\cite{buchbinder1988,buchbinder1992}. It is clear that the source of this problem is the symmetry of $\epsilon^{\alpha}$. Geometrically, it can be understood as $\epsilon^{\alpha}$ taking all possible values on the orbit spanned by $\check{\chi}^{a}$ in parameter space (dashed orbit), as illustrated in FIG. \ref{fig1}.
The key to resolving this issue is to simultaneously fix a point in the parameter-space orbit while demanding condition (\ref{dgp2}).\\
Let, $\check{\chi}^{a}[\epsilon]$ be the coordinates on parameter-space orbit. We demand that the equation,
\begin{eqnarray}
\label{dgpr2}
\chi^{\alpha}[\varphi_{\epsilon}]|_{\check{\chi}^{a}[\epsilon]=\check{\chi}^{a}} = \chi^{\alpha}[\varphi]|_{\check{\chi}^{a}[\epsilon]=\check{\chi}^{a}},
\end{eqnarray}
have unique solution $\delta\epsilon^{\alpha} = 0$. Condition (\ref{dgpr2}) ensures that the change from $\varphi$ to $\varphi_{\epsilon}$ in the field-space orbit happens with respect to a fixed point $\check{\chi}^{a}$ in parameter-space orbit. Hence, eq. (\ref{dgpr2}) is the correct requirement for gauge fixing in the case where $\delta\epsilon^{\alpha}$ has additional symmetry. \\
A more useful form of (\ref{dgpr2}) can be obtained by expressing $\chi^{\alpha}[\varphi]$ as a functional of $\varphi^{i}$, $\epsilon^{\alpha}$ and $\check{\chi}^{a}$ so that,
\begin{eqnarray}
\label{dgpr3}
\chi^{\alpha}[\varphi_{\epsilon}] = \chi^{\alpha}[\varphi,\epsilon,\check{\chi}].
\end{eqnarray}
Substituting in (\ref{dgp2}) and expanding about $\epsilon = 0$ keeping $\varphi$ and $\check{\chi}$ constant gives,
\begin{eqnarray}
\label{dgpr4}
\left(\dfrac{\delta}{\delta\epsilon^{\beta}}\chi^{\alpha}[\varphi,\epsilon ,\check{\chi}]\right)_{\epsilon = 0}\delta\epsilon^{\beta} \equiv Q'^{\alpha}_{\beta}\delta\epsilon^{\beta} = 0 .
\end{eqnarray}
Moreover, expanding both sides of eq. (\ref{dgpr3}) about $\varphi$ results in, 
\begin{eqnarray}
\label{dgpr5}
Q^{\alpha}_{\beta}\delta\epsilon^{\beta} = Q'^{\alpha}_{\beta}\delta\epsilon^{\beta} + \dfrac{\delta\chi^{\alpha}}{\delta\check{\chi}^{a}}\delta\check{\chi}^{a}.
\end{eqnarray}
Using $\delta\check{\chi}^{a} = \check{\chi}^{a}_{,\beta}\delta\epsilon^{\beta}$ in eq. (\ref{dgpr5}), we get a relation between $Q^{\alpha}_{\beta}$ and $Q'^{\alpha}_{\beta}$, 
\begin{eqnarray}
\label{dgpr6}
Q'^{\alpha}_{\beta} = Q^{\alpha}_{\beta} - \dfrac{\delta\chi^{\alpha}}{\delta\check{\chi}^{a}}\check{\chi}^{a}_{,\beta}.
\end{eqnarray}
We would like to make a couple of comments about $Q'^{\alpha}_{\beta}$. Firstly, $Q'^{\alpha}_{\beta}$ defines the functional derivative $\chi^{\alpha}_{,\beta}$, and it can be explicitly checked for theory (\ref{act4}) that $\det Q'^{\alpha}_{\beta} \neq 0$. Secondly, eq. (\ref{dgpr4}) gives a general expression for calculating ghost determinant. Traditional methods for calculating such determinants involve working out integrals of Faddeev-Popov factor, and are thus specific to a particular theory as well as gauge conditions\cite{buchbinder1992, shapiro2016}. In contrast, a remarkable feature of the result (\ref{dgpr4}) is that, in addition to being independent of gauge conditions and any particular theory, it gives geometric meaning to the resolution of degeneracy in such determinants.

Next step towards writing the effective action is to find the appropriate path integral measure, including the Faddeev-Poppov factor (\ref{dgpn1}). We follow the standard procedure of ref. \cite{parker2009}. Full field space volume element can be written in terms of ($\xi^{A}, \epsilon^{\alpha}$) using the length element,
\begin{eqnarray}
\label{dgpr7}
ds^2 = g_{ij}d\varphi^{i}d\varphi^{j} = h_{AB}d\xi^{A}d\xi^{B} + \gamma_{\alpha\beta}d\epsilon^{\alpha}d\epsilon^{\beta}.
\end{eqnarray}
Also, in the parameter space,
\begin{eqnarray}
\label{dgpr8}
\gamma_{\alpha\beta}d\epsilon^{\alpha}d\epsilon^{\beta} = \gamma^{\perp}_{\alpha\beta}d\epsilon_{\perp}^{\alpha}d\epsilon_{\perp}^{\beta} + \check{\gamma}_{ab}d\lambda^{a}d\lambda^{b}.
\end{eqnarray}
Here, $\xi^{A}$ and $\epsilon^{\alpha}$ orthogonal; and so are $d\epsilon_{\perp}^{\alpha}$ and $d\lambda^{a}$ by definition. $h_{AB}$, $\gamma_{\alpha\beta}$, $\gamma^{\perp}_{\alpha\beta}$ and $\check{\gamma}_{ab}$ are the corresponding induced metric on respective coordinates, whereas $g_{ij}$ is the metric on the full field space. The field-space volume element is given by,
\begin{equation}
\label{dgpr9}
\prod_{i}d\varphi^{i}(\det g_{ij})^{1/2} = \prod_{A}d\xi^{A}\prod_{\alpha}d\epsilon^{\alpha} (\det h_{AB}\det \gamma_{\alpha\beta})^{1/2}.
\end{equation}
The technical difficulty we mentioned earlier is due to the non-trivial structure of parameter space, as shown by (\ref{dgpr8}). For a trivial parameter space, where there are no symmetries, one can show that none of the factors in (\ref{dgpr9}) depend on $\epsilon^{\alpha}$\cite{parker2009} and hence $\prod_{\alpha} d\epsilon^{\alpha}$ integrates out. But, for (\ref{dgpr8}) the gauge group volume element is not trivial. So, to determine which factor integrates out of (\ref{dgpr9}), we must calculate the gauge group volume element first. 
We start with parameter space volume element, 
\begin{equation}
\label{dgprr1}
\prod_{\alpha}d\epsilon^{\alpha}(\det \gamma_{\alpha\beta})^{1/2} = \prod_{\alpha}d\epsilon_{\perp}^{\alpha}\prod_{a}d\lambda^{a}(\det \gamma^{\perp}_{\alpha\beta}\det \check{\gamma}_{ab})^{1/2}.
\end{equation}
Following the arguments of \cite{parker2009}, it can be shown that determinants appearing in the right hand side of eq. (\ref{dgprr1}) are independent of $\lambda^{a}$, thereby making the relevant parameter space measure,
\begin{eqnarray}
\label{dgpr10}
\prod_{\alpha}d\epsilon_{\perp}^{\alpha}(\det \gamma^{\perp}_{\alpha\beta})^{1/2}(\det \check{\gamma}_{ab})^{1/2}.
\end{eqnarray}
Now, we introduce the Faddeev-Popov factor,
\begin{eqnarray}
\label{dgp7}
1 = \int \left(\prod_{a}d\check{\chi}^{a}\right)\tilde{\delta}[\check{\chi}^{a}[\epsilon_{\lambda}];0] ,
\end{eqnarray}
so that, the parameter space measure becomes, 
\begin{equation}
\label{dgprr2}
[d\epsilon] = \prod_{\alpha}d\epsilon_{\perp}^{\alpha}\prod_{a}d\check{\chi}^{a}(\det \gamma^{\perp}_{\alpha\beta})^{1/2}(\det \check{\gamma}_{ab})^{1/2}\tilde{\delta}[\check{\chi}^{a}[\epsilon_{\lambda}];0].
\end{equation}
To calculate the Jacobian for transformation from ($\epsilon_{\perp}^{\alpha}, \check{\chi}^{a}$) to $\epsilon^{\alpha}$, we use,
\begin{eqnarray}
\label{dgprr3}
d\check{\chi}^{a} = \check{\chi}^{a}_{,\alpha}d\epsilon^{\alpha} = \check{\chi}^{a}_{,\alpha}d\epsilon_{\perp}^{\alpha} + \check{Q}^{a}_{b}d\lambda^{b},
\end{eqnarray}
where, $\check{Q}^{a}_{b} = \check{\chi}^{a}_{,\alpha}\check{K}^{\alpha}_{b}$. Solving for $d\lambda^{a}$ gives,
\begin{eqnarray}
\label{dgprr4}
d\lambda^{a} = \left(\check{Q}^{-1}\right)^{a}_{b}(d\check{\chi}^{b} - \check{\chi}^{a}_{,\alpha}d\epsilon_{\perp}^{\alpha}).
\end{eqnarray}
Substituting (\ref{dgprr4}) in (\ref{dgpr8}), length element in parameter space is obtained as,
\begin{eqnarray}
\label{dgprr5}
\gamma_{\alpha\beta}d\epsilon^{\alpha}d\epsilon^{\beta} = \gamma^{\perp}_{\alpha\beta}d\epsilon_{\perp}^{\alpha}d\epsilon_{\perp}^{\beta} + \check{\gamma}_{ab}\left(\check{Q}^{-1}\right)^{a}_{c}\left(\check{Q}^{-1}\right)^{b}_{d} \nonumber \\ \times(d\check{\chi}^{c} - \check{\chi}^{c}_{,\alpha}d\epsilon_{\perp}^{\alpha})(d\check{\chi}^{d} - \check{\chi}^{d}_{,\alpha}d\epsilon_{\perp}^{\alpha}).
\end{eqnarray}
The metric in (\ref{dgprr5}) has the form of that in Kaluza-Klein theory\cite{parker2009}, so it is straightforward to read off the relation between volume elements,
\begin{eqnarray}
\label{dgprr6}
(\det\gamma_{\alpha\beta})^{1/2}\prod_{\alpha}d\epsilon^{\alpha} = \left(\prod_{\alpha}d\epsilon_{\perp}^{\alpha}\prod_{a}d\check{\chi}^{a}\right)\times\nonumber \\ (\det \gamma^{\perp}_{\alpha\beta})^{1/2}(\det \check{\gamma}_{ab})^{1/2}(\det\check{Q}^{a}_{b})^{-1}.
\end{eqnarray}
Hence, the gauge group volume is found to be,
\begin{eqnarray}
\label{dgprr7}
[d\epsilon] = \prod_{\alpha}d\epsilon^{\alpha} (\det\gamma_{\alpha\beta})^{1/2} (\det\check{Q}^{a}_{b}[\epsilon])\tilde{\delta}[\check{\chi}^{a}[\epsilon_{\lambda}];0].
\end{eqnarray}
It is safe to say now, that the factor $\prod_{\alpha}d\epsilon^{\alpha}(\det\check{Q}^{a}_{b}[\epsilon])\tilde{\delta}[\check{\chi}^{a}[\epsilon_{\lambda}];0]$ will integrate out of the field space measure. Therefore, we express the field-space measure as,
\begin{eqnarray}
\label{dgprr8}
\prod_{A}d\xi^{A} (\det h_{AB})^{1/2}(\det \gamma_{\alpha\beta})^{1/2} (\det\check{Q}^{a}_{b})^{-1}.
\end{eqnarray}
At this point, it is standard to introduce the Faddeev-Popov factor given by (\ref{dgpn1}), at $\check{\chi}^{a}=0$, and calculate path integral measure by working out Jacobian of coordinate transformations to full field space coordinates $\varphi^{i}$\cite{parker2009}. We need to calculate the field-space measure,
\begin{eqnarray}
\label{dgp9}
[d\varphi] = \int \left(\prod_{A}d\xi^{A}\prod_{\alpha}d\chi^{\alpha}\right)\tilde{\delta}[\chi^{\alpha}[\varphi_{\epsilon}]|_{\check{\chi}^{a}=0};0]\nonumber \\ \times (\det h_{AB} \det \gamma_{\alpha\beta})^{1/2}(\det\check{Q}^{a}_{b})^{-1}.
\end{eqnarray}
Calculation of Jacobian proceeds in a similar way as in eqs. (\ref{dgprr2}) to (\ref{dgprr6}). Using the result (\ref{dgpr6}), 
\begin{eqnarray}
\label{dgpn2}
d\chi^{\alpha} = \chi^{\alpha}_{,A}d\xi^{A} + Q'^{\alpha}_{\beta}d\epsilon^{\beta},
\end{eqnarray}
and solving for $d\epsilon^{\alpha}$ followed by substituting in (\ref{dgpr7}), one obtains the relation, 
\begin{eqnarray}
\label{dgpn3}
\prod_{i}d\varphi^{i}(\det g_{ij})^{1/2}(\det Q'^{\alpha}_{\beta}) = \prod_{A}d\xi^{A}\prod_{\alpha}d\chi^{\alpha} \nonumber \\ \times (\det h_{AB})^{1/2}(\det \gamma_{\alpha\beta})^{1/2}.
\end{eqnarray}
Finally, substituting (\ref{dgpn3}) in (\ref{dgp9}), we obtain the path integral measure for fields $\varphi^{i}$,
\begin{eqnarray}
\label{dgpn5}
[d\varphi] = \int \left(\prod_{i}d\varphi^{i}\right)(\det g_{ij})^{1/2}(\det Q'^{\alpha}_{\beta})\nonumber \\ \times(\det\check{Q}^{a}_{b})^{-1}\tilde{\delta}[\chi^{\alpha}[\varphi_{\epsilon}]|_{\check{\chi}^{a}=0};0].
\end{eqnarray}

\section{Effective Action for rank-2 antisymmetric field}
Due to the properties of covariant effective action approach, the above result (\ref{dgpn5}) is useful to study quantum properties of theories with degeneracy in gauge parameters in curved spacetime, including studies of quantum gravitational corrections\cite{parker2009,saltas2017}. Some interesting studies may include non-minimal models of antisymmetric field with couplings to gravity. Although such studies can be carried out in future, for the purpose of present work we demonstrate the technique as an  application to a massive antisymmetric field in curved spacetime\cite{buchbinder2008,shapiro2016}, as described by Eq. (\ref{act1}) and equivalently, (\ref{act4}). 

The first step is to choose the field space metric. We write the length element in the field space, 
\begin{eqnarray}
\label{eac1}
ds^{2} = \int dv_{x} \left\{g^{\mu\rho}(x)g^{\sigma\nu}(x)dB_{\mu\nu}(x)dB_{\rho\sigma}(x)\right. \nonumber \\ \left. + g^{\mu\nu}(x)dC_{\mu}(x)dC_{\nu}(x)\right\}.
\end{eqnarray}
From (\ref{eac1}), we read off the field space metric components $G_{I(x)J(x^{\prime})}$, taking into account the antisymmetrization of metric component for $B_{\mu\nu}$ fields, 
\begin{eqnarray}
\label{eac2}
G_{B_{\mu\nu}(x)B_{\rho\sigma}(x')} &=& \sqrt{-g(x)} g^{\mu[\rho}(x)g^{\sigma]\nu}(x')\tilde{\delta}(x,x'), \nonumber \\
G_{C_{\mu}(x)C_{\nu}(x')} &=& \sqrt{-g(x)} g^{\mu\nu}(x)\tilde{\delta}(x,x'),
\end{eqnarray}
where $g^{\mu\nu}$ is the spacetime metric and $\tilde{\delta}(x,x')$ is the invariant delta function.
We can calculate the inverse field space metric using the identity 
\begin{equation}
\label{eac3}
\int d^{4}x' G^{I(x)J(x^{\prime})}G_{J(x^{\prime})K(y)} = \delta^{I}_{K}\delta(x,y),
\end{equation}
which gives,
\begin{eqnarray}
\label{eac4}
G^{B_{\mu\nu}(x)B_{\rho\sigma}(x')} &=& \dfrac{1}{-g(x)}g_{\mu[\rho}(x)g_{\sigma]\nu}(x')\delta(x,x'), \nonumber \\
G^{C_{\mu}(x)C_{\nu}(x')} &=& \dfrac{1}{-g(x)} g_{\mu\nu}(x)\delta(x,x').
\end{eqnarray}
Since, in (\ref{eac2}) and (\ref{eac4}), there is no dependence on the fields, the field space Christoffel connections will vanish, 
\begin{equation}
\label{eac5}
\Gamma^{i}_{jk} = 0 \quad \forall \quad  i,j,k \in (B_{\mu\nu}, C_{\mu}).
\end{equation}
Note that the Christoffel connections will be nonzero if we choose to quantize gravity as well. For the present case, however, we treat gravity as a classical field so that field space only has components of  $B$ and $C$ fields and the connections vanish. Another case where Christoffel connections would be nonzero is if we choose a different parametrization for the fields. This corresponds to choosing another coordinate system in field space. \\
Next, we find the gauge generators $K^{i}_{\alpha}$ from the relation
\begin{eqnarray}
\label{eac6}
\delta\varphi^{I} = \int d^{4}x' K^{\varphi^{I}(x)}_{\bar{\alpha}(x')}\delta\epsilon^{\bar{\alpha}(x')},
\end{eqnarray}
where $\varphi^{I(x)} = \{B_{\mu\nu}(x), C_{\mu}(x)\}$ and $\delta\epsilon^{\bar{\alpha}(x')} = \{\xi_{\mu}(x), \Lambda(x)\}$. The generators can be read off from eqs. (\ref{act5}) and (\ref{act6}), 
\begin{eqnarray}
\label{eac7}
K^{B_{\mu\nu}(x)}_{\xi_{\rho}(x')} &=& (\nabla_{\mu}\delta^{\rho}_{\nu} - \nabla_{\nu}\delta^{\rho}_{\mu})\tilde{\delta}(x,x'), \nonumber \\
K^{C_{\mu}(x)}_{\xi_{\rho}(x')} &=& -m\delta^{\rho}_{\mu}\tilde{\delta}(x,x'), \\
K^{C_{\mu}(x)}_{\Lambda(x')} &=& \nabla_{\mu}\tilde{\delta}(x,x'). \nonumber
\end{eqnarray}
For future use, we also calculate $K_{\alpha i}$ below, using the condensed notation identity $K_{\alpha i} = g_{ij}K^{j}_{\alpha} = \int d^{4}y G_{I(x)J(y)}K^{J(y)}_{\bar{\alpha}(x')}$:
\begin{eqnarray}
\label{eac7b}
K_{\xi_{\mu}(x')B_{\nu\rho}(x)} &=& \sqrt{-g(x)}(\nabla^{\nu}\delta^{\rho\mu}-\nabla^{\rho}\delta^{\nu\mu})\tilde{\delta}(x,x'), \nonumber \\
K_{\Lambda(x')C_{\mu}(x)} &=& \sqrt{-g(x)} \nabla^{\mu} \tilde{\delta}(x,x'), \\
K_{\xi_{\rho}(x')C_{\mu}(x)} &=& -m \sqrt{-g(x)} \delta^{\mu\rho} \tilde{\delta}(x,x'). \nonumber
\end{eqnarray}
Keeping in mind the geometrical interpretation of gauge fixing, we understand that for effective action only a sub-space of field space constrained by the gauge condition will be integrated over, and hence the metric with which covariant derivatives and connections are calculated in field space is not the full field space metric $G_{I(x)J(x')}$, but that which describes the space of transverse fields $\xi^{A}$. 
However, for the present example, since $K^{i}_{\alpha}$ does not have any dependence on the fields $B_{\mu\nu}$ and $C_{\mu}$, it turns out the induced connection vanishes for all combinations of $B_{\mu\nu}$ and $C_{\mu}$:
\begin{equation}
\label{eac14}
\tilde{\Gamma}^{m}_{ij} = 0.
\end{equation}
This simplifies the present problem a lot. In order to find the 1-loop corrections, covariant derivatives of the action become merely simple functional derivatives.\\
The covariant field-space intervals also reduce to flat geodetic intervals due to vanishing connections, 
\begin{equation}
\label{eac15}
\sigma^{i} = -\eta^{i} = \varphi^{i}_{*} - \varphi^{i}.
\end{equation}

Next, we use the result (\ref{dgpn5}) to write the gauge-fixed measure. For convenience, we choose for $B_{\mu\nu}$ the gauge condition,
\begin{eqnarray}
\label{eac20}
\chi_{\xi_{\nu}} = \nabla^{\mu}B_{\mu\nu} + m C_{\nu},
\end{eqnarray}
and for gauge parameters,
\begin{eqnarray}
\label{eacf5}
\check{\chi}_{\psi} = \nabla^{\mu}\xi_{\mu} - m\Lambda
\end{eqnarray}
This results in the action (\ref{act4}) being appended by a gauge fixing term,
\begin{eqnarray}
\label{eacf1}
-\frac{1}{2}(\chi_{\xi_{\nu}})^{2} = -\frac{1}{2}\nabla^{\mu}B_{\mu\nu}\nabla_{\rho}B^{\rho\nu}  - \frac{1}{2}m^{2}C_{\nu}C^{\nu} \nonumber \\ - \frac{1}{2}m C_{\nu}\nabla_{\mu}B^{\mu\nu}.
\end{eqnarray}
The second term on the right hand side of eq. (\ref{eacf1}) induces soft breaking of gauge symmetry in field $C_{\nu}$. As pointed out earlier, one has to apply the St\"{u}ckelberg procedure of sec. \ref{sec2} to restore gauge symmetry. As a result, a second St\"{u}ckelberg field $\Phi$ is introduced so that,
\begin{eqnarray}
\label{eacf2}
- \frac{1}{2}m^{2}C_{\nu}C^{\nu} \ \longrightarrow \ - \frac{1}{2}m^{2}\left(C^{\nu} + \frac{1}{m}\nabla^{\nu}\Phi\right)^{2}.
\end{eqnarray}
Hence, an appropriate gauge condition for $C_{\mu}$ is,
\begin{eqnarray}
\label{eacf3}
\chi_{\Lambda} = \nabla^{\mu}C_{\mu} + m\Phi .
\end{eqnarray}
In order to calculate the components of $Q'^{\alpha}_{\beta}$, we use eq. (\ref{dgpr3}) to write, 
\begin{eqnarray}
\label{eacf4}
\chi_{\xi_{\nu}} = \nabla^{\mu}B_{\mu\nu} + m C_{\nu} + \Box_{x}\xi_{\nu} - [\nabla^{\mu},\nabla_{\nu}]\xi_{\mu} \nonumber \\ - m^{2}\xi_{\nu} - \nabla_{\nu}\check{\chi}_{\psi}
\end{eqnarray}
From the definition (\ref{dgpr4}), we find,
\begin{eqnarray}
\label{eacf6}
Q'^{\xi_{\mu}(x)}_{\xi_{\nu}(x')} &=& (\Box_{x}\delta_{\nu\mu} - [\nabla^{\alpha},\nabla_{\nu}]\delta_{\alpha\mu} - m^{2}\delta_{\nu\mu})\tilde{\delta}(x,x') \nonumber \\ 
&\equiv & (\Box_{1} - m^{2}\delta_{\nu\mu})\tilde{\delta}(x,x').
\end{eqnarray}
Calculation of $Q'^{\Lambda(x)}_{\Lambda(x')}$ is straightforward, and yields,
\begin{eqnarray}
\label{eac23}
Q'^{\Lambda(x)}_{\Lambda(x')} &=& (\Box_{x}-m^2)\tilde{\delta}(x,x').  
\end{eqnarray}
A similar calculation results in $\check{Q}^{\psi(x)}_{\psi(x')} = (\Box_{x} - m^{2})\times$ $\tilde{\delta}(x,x')$.
With all the ingredients in place, we obtain the effective action by substituting eqs. (\ref{eac20}), (\ref{eacf3}), and (\ref{eac23}) in (\ref{dgpn5}) and working in spacetime (uncondensed) coordinates, 
\begin{eqnarray}
\label{eac25}
\exp(i\Gamma[\bar{B},\bar{C}]) = \int\prod_{\mu}dC_{\mu}\prod_{\rho\sigma}dB_{\rho\sigma}\prod_{x}d\Phi\times\nonumber \\ 
\delta[\chi_{\Lambda(x)};0]\delta[\chi_{\xi_{\mu}(x)};0]\det(\Box_{1} - m^{2})\times \nonumber \\  
 \exp\left\{i \int d v_{x} \left(-\dfrac{1}{12}F^{\mu\nu\lambda}[B]F_{\mu\nu\lambda}[B] \right.\right. \nonumber \\ \left. \left.
-\dfrac{1}{4}m^{2}(B^{\mu\nu} + \frac{1}{m}H^{\mu\nu}[C])^{2}\right) \right. \nonumber \\ \left. + (\bar{B}_{\mu\nu}-B_{\mu\nu})\dfrac{\delta}{\delta \bar{B}_{\mu\nu}}\Gamma[\bar{B},\bar{C}] \right. \nonumber \\ \left. + (\bar{C}_{\mu}-C_{\mu})\dfrac{\delta}{\delta \bar{C}_{\mu}}\Gamma[\bar{B},\bar{C}] \right\}.
\end{eqnarray}
We have ignored the field-space metric determinant here, because it does not affect the result apart from raising and lowering indices inside determinants. As usual, the Dirac $\delta$-distributions in (\ref{eac25}) give rise to the gauge-fixed action, 
\begin{eqnarray}
S_{GF} =  \int d v_{x} \Bigg(-\dfrac{1}{12}F^{\mu\nu\lambda}[B]F_{\mu\nu\lambda}[B]  \nonumber \\  - \dfrac{1}{4}m^{2}(B^{\mu\nu} + \frac{1}{m}H^{\mu\nu}[C])^{2} \nonumber \\ -\frac{1}{2}(\chi_{\xi_{\mu}(x)})^{2} - \frac{1}{2}(\chi_{\Lambda(x)})^{2}\Bigg),
\end{eqnarray}
which can be cast into the form,
\begin{eqnarray}
\label{eacr1}
S_{GF} =  \int d v_{x} \Bigg(\frac{1}{4}B^{\mu\nu}\Box_{2}B_{\mu\nu} - \frac{1}{4}m^{2}B^{\mu\nu}B_{\mu\nu}\nonumber \\ + \frac{1}{2}C^{\mu}\Box_{1}C_{\mu} - \frac{1}{2}m^{2}C^{\mu}C_{\mu} \nonumber \\ + \frac{1}{2}\Phi(\Box_{x}-m^2)\Phi\Bigg),
\end{eqnarray}
where, 
\begin{eqnarray}
\Box_{2}B_{\mu\nu} = \Box_{x}B_{\mu\nu} - [\nabla^{\rho},\nabla_{\nu}]B_{\mu\rho} - [\nabla^{\rho},\nabla_{\mu}]B_{\rho\nu} .
\end{eqnarray}
Substituting eq. (\ref{eacr1}) in (\ref{eac25}), one obtains the effective action as,
\begin{eqnarray}
\label{eac28}
\Gamma &=& S[\bar{B},\bar{C}] + \hbar\dfrac{i}{2}\left(\ln\det(\Box_{2} - m^{2}) \right. \nonumber \\ && \left. - \ln\det(\Box_{1} - m^{2}) + \ln\det(\Box_{x}-m^2)\right).
\end{eqnarray}
For the present case of free theory (\ref{act4}), there are no contributions at higher loop orders.\\ The result (\ref{eac28}) is in line with that obtained earlier in \cite{buchbinder2008}, and later confirmed in \cite{shapiro2016}.

\section{Conclusion}
We quantized a massive rank-2 antisymmetric field by finding a general path integral measure for theories with degeneracy in gauge parameter, using DeWitt-Vilkovisky's covariant effective action approach. In the process, we arrived at a simple resolution to the problem of dealing with additional symmetries of gauge parameter, through a geometric understanding of gauge-fixing. 
In particular, the ghost determinant calculation receives a simple general geometric meaning, and generalizes traditional methods\cite{buchbinder1992,shapiro2016} which are specific to a particular theory and gauge condition.

For the simple case of free theory (\ref{act4}), where gravity is classical, we find that the covariant effective action is identical to that obtained in earlier works\cite{buchbinder2008,shapiro2016}, upto a difference in sign of $m^{2}$ due to corresponding sign of potential. More applications of this formalism lie in the study of gravitational corrections to models of antisymmetric tensor fields. For instance, a rather simple generalization of the result (\ref{eac28}) is for the nonminimal model considered in \cite{altschul2010} with a coupling term $\zeta R B_{\mu\nu}B^{\mu\nu}$, which can be absorbed in the mass term with $m\longrightarrow m - \zeta R$. 

\begin{acknowledgments}
This work was partially funded by DST grant no. SERB/PHY/2017041.
\end{acknowledgments}

\bibliography{ref}

\end{document}